\documentclass[12pt]{article}

\newsavebox{\ns}
\newsavebox{\dbrane}
\newsavebox{\dbshort}

\usepackage{epsfig}
\usepackage{amssymb}

\def\be{\begin{eqnarray}}
\def\ee{\end{eqnarray}}

\newcommand{\nn}{\nonumber}
\newcommand\para{\paragraph{}}
\newcommand{\ft}[2]{{\textstyle\frac{#1}{#2}}}
\newcommand{\eqn}[1]{(\ref{#1})}

\def\Dslash{\,\,{\raise.15ex\hbox{/}\mkern-12mu D}}
\def\Dbarslash{\,\,{\raise.15ex\hbox{/}\mkern-12mu {\bar D}}}
\def\delslash{\,\,{\raise.15ex\hbox{/}\mkern-9mu \partial}}
\def\delbarslash{\,\,{\raise.15ex\hbox{/}\mkern-9mu {\bar\partial}}}
\def\pslash{\,\,{\raise.15ex\hbox{/}\mkern-9mu p}}
\def\calDslash{\,\,{\raise.15ex\hbox{/}\mkern-12mu {\cal D}}}

\def\be{\begin{eqnarray}}
\def\ee{\end{eqnarray}}
\def\tr{{\rm Tr}}

\def\D{{\cal D}}
\def\vac{|\,0\rangle}
\def\nn{\nonumber}

\def\bR{\mathbb{R}}
\def\bC{\mathbb{C}}
\def\bH{\mathbb{H}}
\def\bO{\mathbb{O}}

\def\bP{\mathbb{P}}
\def\bK{\mathbb{K}}

\begin{document}
\pagestyle{plain}
\setcounter{page}{1}
\newcounter{bean}
\baselineskip16pt

\begin{titlepage}

\begin{center}
\today
\hfill {.}\\

\vskip 1.5 cm {\large \bf The Berry Phase of D0-Branes}
\vskip 1cm
{Chris Pedder, Julian Sonner and David Tong}\\
\vskip 1cm {\sl Department of Applied Mathematics and Theoretical
Physics, \\ University of Cambridge, UK}

\end{center}

\vskip 0.5 cm
\begin{abstract}
We study $SU(2)$ Yang-Mills quantum mechanics with $N=2,4,8$ and
$16$ supercharges. This describes the non-relativistic dynamics of
a pair of D0-branes moving in $d=3,4,6$ and $10$ spacetime
dimensions respectively. We show that as the D0-branes orbit,
states undergo a Berry holonomy described by the four Hopf maps. For the $N=2$
theory, the associated Hopf map is the ${\bf Z}_2$ M\"obius bundle
and its effect is to turn the D0-branes into anyons with exchange
statistics $e^{i\pi/2}$. For the $N=4,8$ and $16$ theories, the
Hopf maps give rise to Berry connections that are familiar to
physicists: the $U(1)$ Dirac monopole; the $SU(2)$ Yang monopole;
and the $SO(8)$ octonionic monopole.

\end{abstract}

\end{titlepage}

\subsection*{1. Introduction}

Many years ago, Kugo and Townsend \cite{kugo} pointed out a
relationship between  supersymmetric field theories with $N=2,4,8$
and $16$ supercharges and the four normed division algebras
$\bK\cong \bR$, $\bC$, $\bH$ and $\bO$. The key observation is
algebraic. Theories with $N=2,4,8$ and $16$ supercharges naturally
live in $d=3,4,6$ and $10$ spacetime dimensions respectively. The
Lorentz Lie algebra in these dimensions is isomorphic to the
algebra of $2\times 2$ Hermitian matrices with vanishing trace and
elements in $\bK$,
\be sl(2;\bK) \cong so(d-1,1) \label{the}\ee
This allows us to express a spinor in $d$ dimensions as a
2-component $\bK$-vector, generalizing the well-known result for
$d=4$. This construction was elaborated upon in \cite{jon}.

\para
For theories with $N=2,4$ and $8$ supercharges, the relationship
to the division algebra $\bK$ also manifests itself in more
physical and dynamical matters. This includes familiar features of
supersymmetric theories, such as the holomorphy of the
superpotential and the hyperK\"ahler/quaternionic structure of
Calabi-Yau moduli spaces. However, so far the tantalizing idea
that an octonionic structure underlies theories with 16
supercharges has not led to major insight about quantum dynamics.

\para
The purpose of this short note is to point out that the
isomorphism \eqn{the} has a simple, physical consequence in
the framework of supersymmetric quantum mechanics. We focus on
$SU(2)$ gauged quantum mechanics. The theory with $N$ supercharges
can be thought of as the dimensional reduction of the minimal
super Yang-Mills theory in $d=3,4,6$ and 10 dimensions, and
describes the non-relativistic dynamics of two D0-branes moving in
$d-1=N/2+1$ spatial dimensions. We show that the states of a pair
of orbiting D0-branes undergo a holonomy described by the Hopf map
associated to the division algebra $\bR$, $\bC$, $\bH$ and $\bO$.

\para
The four Hopf maps take $S^{N-1}\rightarrow S^{N/2}$:
\be \begin{array}{c} S^{1} \\ \ \,\Bigg{\downarrow} {}^{S^0\cong Z_2} \\
S^1
\end{array}\ \ \ \ \begin{array}{c} S^{3} \\ \ \,\Bigg{\downarrow} {}^{S^1} \\
S^2
\end{array}\ \ \ \ \begin{array}{c} S^{7} \\ \ \,\Bigg{\downarrow} {}^{S^3} \\
S^4
\end{array}\ \ \ \ \begin{array}{c} S^{15} \\ \ \,\Bigg{\downarrow} {}^{S^7} \\
S^8
\end{array}\nn\ee
%
%
%
The case of $N=2$ supercharges is special. The map ${\bf
Z}_2\hookrightarrow S^1\rightarrow S^1$, sometimes known as the
zeroth Hopf map, describes the M\"obius bundle. We will show in
Section 3.1 that the ground state of the associated quantum
mechanics undergoes a discrete ${\bf Z}_2$ holonomy, changing by a
sign as D0-branes orbit. Since a single orbit is equivalent to two
exchanges, this implies that the D0-branes in $N=2$ Yang-Mills
quantum mechanics are anyons: they obey exchange statistics
$e^{\pm i \pi/2}=\pm i$. As we review in Section 3, the remaining
three Hopf maps induce connections over the base space $S^{d-2}$
which subsequently arise as Berry connections in the quantum
mechanics. Each is familiar to physicists: they are the $U(1)$
Dirac monopole, the $SU(2)$ Yang monopole \cite{yang}, and the
$SO(8)$ octonionic monopole \cite{grossman,tchrak}.

\para
This quartet of Hopf bundles are the fab four of geometrical
phases. The discrete ${\bf Z}_2$ holonomy was one of the earliest
known Berry phases \cite{chemist}, while the Dirac monopole
appeared as an example in Berry's original paper \cite{berry}.
Both the Yang monopole and the octonionic monopole have also
arisen in various contexts. The former was first introduced as a
non-Abelian Berry phase in \cite{avron,feher}, and has subsequently
appeared in several condensed matter systems
\cite{bigzhang,middlezhang,smallzhang}, with various properties
explored in \cite{levay}. The $SO(8)$ octonionic monopole appeared
previously in association with the eight-dimensional quantum Hall
effect \cite{bern}. In this paper we will see how all these
connections arise as (non-Abelian) Berry phases in Yang-Mills
quantum mechanics with $N$ supercharges.

\para
The present paper can be thought of as a follow-up to our recent
work \cite{us,second} exploring the Berry phase that emerges in
various supersymmetric quantum mechanics. In \cite{us}, we showed
that certain $N=4$  quantum systems naturally give rise to the
Dirac monopole and associated constructs such as the smooth 't
Hooft-Polyakov monopole. In \cite{second}, we studied a quantum
theory with $N=8$ supercharges and showed that a deformation of
the Yang-monopole arises as the spin connection of a dual
gravitational background. The Yang monopole and octonionic monopole 
also appeared in a related context in \cite{usp}.

\subsection*{2. D0-Brane Dynamics}

In this paper we study the $SU(2)$ super-Yang-Mills quantum
mechanics with $N=2,4,8$ and $16$ supercharges. The Lagrangians
are the dimensional reduction of minimal super Yang-Mills in
$d=3,4,8$ and $10$ dimensions respectively, and each takes the
form
\be \label{lag} L = \frac{1}{2g^2}\,\tr \left( \left( \D_0 X_i
\right)^2 + \sum_{i<j}\left[ X_i , X_j \right]^2 +i \bar\psi\D_0
\psi + \bar\psi \Gamma^i \left[ X_i , \psi \right]
\right)\,, \ee %
where $X_i$ and $\psi_\alpha$ are both $su(2)$ valued fields. The
inverse coupling constant $1/g^2$ is the mass of the D0-brane.
Lagrangians with different amounts of supersymmetry are
distinguished by the number of scalar fields $X_i$, and the
structure of the Grassmann parameters $\psi$. The scalar index
runs over $i=1,\ldots, d-1$ and the Lagrangian describes the
non-relativistic relative motion of a pair of D0-branes moving in
$d-1=2,3,5$ and $9$ spatial dimensions\footnote{More precisely,
the theory with $N=16$ supercharges  describes the dynamics of a
pair of D0-branes in flat ten-dimensional spacetime. Lagrangians
with less supersymmetry describe the dynamics of a pair of
fractional D0-branes, trapped to lie at a suitable singularity in
a K3, a CY 3-fold, or a $G_2$ holonomy manifold.}. The matrices
$\Gamma^i$ satisfy the $d-1$ dimensional Euclidean Clifford
algebra: $\{\Gamma^i,\Gamma^j\}=2\delta^{ij}$. The spinors are
real or complex, depending on the possible representations of the
Clifford algebra:
\begin{itemize}
\item $N=2$: For $d-1=2$, we may choose the real Pauli matrices
$\Gamma^1=\sigma^1$, $\Gamma^2=\sigma^3$. The spinor
$\psi_\alpha$, with $\alpha=1,2$, is also real.
\item $N=4$: For $d-1=3$, there are only complex representations
of the Clifford algebra. We take the Pauli matrices
$\Gamma^i=\sigma^i$. The spinor $\psi_\alpha$, with $\alpha=1,2$,
is complex.
\item $N=8$: The Clifford algebra in $d-1=5$ dimensions is again
complex. The complex spinor $\psi_\alpha$ has $\alpha=1,\ldots,4$.
\item $N=16$: The $d-1=9$ Clifford algebra admits a real
representation. The real spinor $\psi_\alpha$ has
$\alpha=1,\ldots,16$.
\end{itemize}

\subsubsection*{2.1 The Born-Oppenheimer Approximation}

We work in the Born-Oppenheimer approximation, considering
well-localized wavepackets describing D0-branes with separation
$\hat{X}^i$. To this end, we expand around the classical
background
\be \langle X^i\rangle = \ft12 \hat{X}^i\sigma^3 \label{bo}\ee
and study the effects of high-frequency modes of open strings
stretched between the D0-branes. Our results are applicable in the
weak coupling limit $\hat{X}^3\gg g^2$. This typically excludes
the ground state of the full system, for which the wavefunction is
localized close to the origin $\hat{X} \sim g^{2/3}$.

\para
The background \eqn{bo} breaks the gauge symmetry to the Cartan
subalgebra: $SU(2)\rightarrow U(1)$. The remnant ${\bf Z}_2$ Weyl
group acts as $\hat{X}^i\rightarrow -\hat{X}^i$. The configuration
space of the D0-branes is thus ${\bf R}^d/{\bf Z}_2$,  reflecting
the indistinguishability of the particles.  We expand,
\be \label{bmasses} X^i = \langle X^i\rangle + \ft12 x^i_m
\sigma^m \ee
The bosonic commutator term in \eqn{lag} provides oscillation
frequencies for the fluctuations $x^i_m$. The fluctuations
$x^i_3$, lying in the unbroken $U(1)$, are zero modes and describe
the relative motion of the D0-brane pair in $\bR^{d-1}$. In
contrast, the $m=1,2$ components, lying in the broken part of the
gauge group, pick up non-zero frequencies and describe the
excitation of open strings stretched between the D0-branes. We
form the complex combination
\be z^i=x^i_1+ix^i_2\ ,\ \ \ \ i=1,\ldots, d-1\ee
which has charge $+1$ under the unbroken $U(1)$ gauge symmetry.
Introducing the complex conjugate momentum $\pi_i$, the leading
order free Hamiltonian for these off-diagonal bosonic modes is
given by,
\be H_B = \frac{1}{2g^2}\left[ {\cal P}_{ij}\left( \pi_i
\bar\pi_j+ \hat X^2 z_i\bar z_j \right) + \ldots
\right] \ee
where $\ldots$ are interaction parts of the Hamiltonian which are
perturbations of order $g/\hat{X}^{1/3}$. The Hamiltonian includes
the projection operator
\be {\cal P}_{ij} = \delta_{ij}-\frac{\hat X^i \hat X^j}{\hat
X^2}\label{bproj}\ee
and therefore describes only $d-2$ complex harmonic oscillators.
These oscillators can be thought of as the transverse excitations
of a string stretched between the D0-branes. The remaining degree
of freedom falls victim to the broken gauge symmetry on the
D0-branes; it is analogous to the scalar that is eaten by the
Higgs mechanism in higher dimensions.

\para
We may also expand the fermions around the background \eqn{bo}.
Once again, the fermions $(\psi_\alpha)_3$, lying in the Cartan
subalgebra, provide zero modes. Upon quantization, these fill out
a $2^{N/2}$ dimensional multiplet of states whose degeneracy is
split only by interaction terms. In contrast, the off-diagonal
components $(\psi_\alpha)_1$ and $(\psi_\alpha)_2$ have non-zero
frequencies. Here it is convenient to differentiate between the
$N=2,16$ cases, which have real fermions and the $N=4,8$ cases,
which have complex fermions. In the former case, we introduce the
complexified Grassmann parameter,
\be \label{eq:complexspinor} \Psi=\psi_1 + i \psi_2\,. \ee
in terms of which the fermionic Hamiltonian is written as
\be H_F^{N=2,16} = \frac{1}{2g^2}\left[\Psi^\dagger \left( \hat X
\cdot \Gamma \right)\Psi + \ldots\right]\label{hf216}\ee
The $N=4,8$ cases have complex spinors from the outset. We may now
form two linearly independent complex combinations
\be \Psi =  \frac{1}{\sqrt{2}} \left(  \psi_1 + i \psi_2 \right)\
\ \ ,\ \ \  \tilde{\Psi} =  \frac{1}{\sqrt{2}} \left( \psi_1 - i
\psi_2 \right) \ee
with respective charges $+1$ and $-1$ under the unbroken $U(1)$
gauge group. The fermionic part of the Hamiltonian is now given by
\be H_F^{N=4,8}=\frac{1}{2g^2}\left[\Psi^\dagger \left(\hat X
\cdot \Gamma \right)\Psi - \tilde{\Psi}^\dagger \left(  \hat X
\cdot \Gamma \right)\tilde{\Psi}+\ldots\right] \label{hf48}\ee
To summarize, the free part of the Hamiltonian $H=H_B+H_F$ for the
massive oscillators contains $N/2$ complex scalars and $N$ complex
Grassmann parameters. The interaction Hamiltonian is of order
$g/\hat{X}^{1/3}$.

\subsubsection*{2.2 Quantization and the Hilbert Space}

In the Born-Oppenheimer approximation, we treat the massive
oscillator states $z_i$, $\Psi_\alpha$ and $\tilde{\Psi}_\alpha$
quantum mechanically in the classical background $\hat{X}$. The
zero frequency modes $\hat{X}^i$ and $(\psi_\alpha)_3$ are
quantized subsequently. The approximation holds as long as
$\hat{X}\gg g^{1/3}$ ensuring that we keep a separation of scales,
meaning that the wavefunction for $\hat{X}$ should not have
significant support near the origin where the two D0-branes
approach. For example, this will be the case for excited states of
orbiting D0-branes that carry large angular momentum. In the
following section we will study the holonomy of the excited states
of the massive oscillators as the two D0-branes orbit. We first
briefly describe the Hilbert space of these states.

\para
Working in the regime $\hat{X}\gg g^{1/3}$, we may restrict
attention to the free theory. The $d-2$ massive complex scalars
have ground state energy  $E_B=(d-2)\hat{X}/2g^2$. One may
construct the Hilbert space by acting with (suitably projected)
creation operators ${\cal P}_{ij}\,a^\dagger_j$ and ${\cal
P}_{ij}\,\bar{a}^\dagger_j$, where
\be a^\dagger_i=\frac{(\hat{X}z_i
-i\bar{\pi}_i)}{\sqrt{2\hat{X}}}\ \ \ ,\ \ \ \bar{a}_i^\dagger =
\frac{(\hat{X} \bar{z}_i-i\pi_i)}{\sqrt{2\hat{X}}} \ee
create quanta of charge $+1$ and $-1$ respectively under the
unbroken $U(1)$ gauge symmetry. Canonical quantization for
fermions gives the brackets $\{ \Psi_\alpha,\Psi_{\beta}^\dagger
\} = \delta_{\alpha\beta}$ and
$\{\tilde{\Psi}_\alpha,\tilde{\Psi}_{\beta}^\dagger \} =
\delta_{\alpha\beta}$. We build the fermionic Hilbert space ${\cal
H}_F$ by picking a reference state $\vac$, satisfying
\be \Psi_\alpha\vac=\tilde{\Psi}_\alpha\vac=0\label{ref}\ee
We may then act  upon $\vac$ with $\Psi^\dagger$ in the $N=2,16$
theories, and $\Psi^\dagger$ and $\tilde{\Psi}^\dagger$ in the
$N=4,8$ theories, to construct a Hilbert space of dimension
$\dim({\cal H}_F)=2^N$.

\para
The free fermionic Hamiltonians \eqn{hf216} and \eqn{hf48} have a
unique ground state $|\Omega\rangle$ with energy
$E_F=-N\hat{X}/4g^2$. This ensures that the ground state energy of
the full theory is $E_0=E_B+E_F=0$. The fermionic ground state
always lies in the sector with half of the fermions excited. To
describe it, we first introduce the fermionic projection operators
\be P_\pm = \frac{1}{2}\left( 1 \pm \frac{\hat X\cdot
\Gamma}{|\hat X|}\right)\label{fproj}\ee
Then, for the $N=2,16$ theories, the (un-normalized) ground state
is given schematically by,
\be\label{eq:ground} |\Omega\rangle = \left(P_-
\Psi^\dagger\right)^{N/2} \vac \label{vac1}\ee
while, for the $N=4,8$ theories, the (un-normalized) ground state
is given by
\be |\Omega\rangle = \left(P_- \Psi^\dagger\right)^{N/4} \left(P_+
\tilde{\Psi}^\dagger\right)^{N/4}   \vac \label{vac2}\ee
Note that $|\Omega\rangle$ is the ground state for the high
frequency modes in the Born-Oppenheimer approximation. The
question of whether a normalizable ground state of the full theory
exists is more subtle, and irrelevant for our considerations. (It
does for $N=16$, but is expected not to for $N=2,4,8$).

\para
The physical Hilbert space of the theory is subject to Gauss' law
for the unbroken $U(1)\subset SU(2)$ gauge symmetry which ensures
that all physical states are gauge neutral. The vacuum
$|\Omega\rangle$ is assigned charge zero under the $U(1)$ and so
survives Gauss' purge. Other states in the fermionic Hilbert space
arise by acting on $|\Omega\rangle$ with combinations of the
charge $+1$ states $P_-\Psi$ and $P_-\tilde{\Psi}^\dagger$ and the
charge $-1$ states $P_+\Psi^\dagger$ and $P_+\tilde{\Psi}$. In
each case, one may construct a state in the physical Hilbert space
by dressing the fermionic state with appropriate powers of the
charged bosonic operator ${\cal P}_{ij}a_j^\dagger$ and ${\cal
P}_{ij}\bar{a}_j^\dagger$.

\subsection*{3. Berry Phase and Hopf Maps}

In the previous section we constructed the Hilbert space over each
background separation $\hat{X}^i$. In this section we are
interested in how these Hilbert spaces evolve as the D0-branes
orbit, and $\hat{X}^i$ traces a closed path $\Gamma$ in
configuration space. The dynamical phase of a set of degenerate
states $|\phi_a\rangle$ is accompanied by a (possibly non-Abelian)
Berry holonomy, described by
\be |\phi_a\rangle \longrightarrow P \exp\left(-i \oint_\Gamma
{A}_{ab}\cdot \hat{X} \right)|\phi_b\rangle \ \ \ {\rm with}\ \ \
(A_i)_{ab}=i\langle\phi_b|\frac{\partial}{\partial\hat{X}^i}|\phi_a\rangle\ee
We will see that for certain states in quantum mechanics with $N$ supercharges, the
Berry connection $A$ is given by the associated Hopf map. We start
by reviewing the the Hopf maps and connections in more detail.

\para
The Hopf map from $S^{N-1}\rightarrow S^{N/2}$ is defined in the
following manner. One starts with a commuting spinor
$\chi_\alpha$, of the type described in Section 2: real for
$N=2,16$ and complex for $N=4,8$. Imposing the normalization
condition $\chi^\dagger \chi =1$ ensures that $\chi$ defines a
point on $S^{N-1}$. The map to the base manifold $S^{N/2}$ is then
given by the bi-linear form
\be n^i = \chi^\dagger \Gamma^i \chi \ee
where $\Gamma^i$ obey the $SO(d-1)$ Clifford algebra. One may
check that $n^in^i=1$ and hence $n^i$ defines a point on
$S^{N/2}$.

\para
It is illustrative to stress the connection to the division
algebras by reformulating the Hopf maps over the algebra $\bK$
using the isomorphism \eqn{the}. (For more details see, for example,
\cite{baez}). The total manifold $S^{N-1}$ can be thought of as
arising from the pair $(q_1,q_2)\in \bK^2$, subject to the
constraint $|q_1|^2 + |q_2|^2=1$. From this pair we can define the
vector $n^i$, $i=1\ldots,d-1$,
\be n^i = q^\dagger \Gamma^i q\ee
where $\Gamma^i$ still satisfy the $SO(d-1)$ Clifford algebra,
hence justifying their name, but can now be thought of as basis
elements of $\Gamma^i \in sl(2;\bK)$, defined by
\be \Gamma^i=\left(\begin{array}{cc} 0 & e_i \\ e_i^\star & 0
\end{array}\right)\ \  \ i=1,\ldots , N/2\ \ \ ,\ \ \
\Gamma^{d-1}=\left(\begin{array}{cc} 1 & 0 \\ 0 & -1
\end{array}\right)\label{gam}\ee
where $e_i$ are the generators of the division algebra $\bK$. The
fact that these matrices obey the Clifford algebra reflects  the
isomorphism \eqn{the}. One can check once again that $n^in^i=1$,
ensuring that this defines a map to the base manifold
$S^{N/2}\cong \bK\bP^1$.

\para
The zeroth Hopf map associated to $\bR$ is the M\"obius bundle
${\bf Z}_2\hookrightarrow S^1\rightarrow
\bR\bP^1$. We will shortly see how this
arises in the theory with $N=2$ supercharges: the holonomy of the
ground state is a minus sign whose role is to render the D0-branes
anyonic. The remaining Hopf maps, associated to $\bC$, $\bH$ and
$\bO$, each define a connection $A$ over the base space $S^{N/2}$,
which describes how the base manifold is twisted inside the total
space. These connections are given in terms of the projection
operators $P_\pm$ defined in \eqn{fproj}, where we identify
$\hat{X}^i\equiv \hat{X}n^i$. Let $\lambda_a$ be the non-vanishing
orthonormal eigenvectors of $P_-$: i.e. $P_-\lambda_a=\lambda_a$.
The Hopf connection is defined by,
\be A_{ab} = i\lambda_b^\dagger\, d\lambda_a\label{hopfcon}\ee
For $N=4,8$ and $16$, the maximum value of the index $a$ is
$1,2,8$ and $A_{ab}$ is therefore a $U(1)$, $U(2)$ and $SO(8)$
connection respectively. In the following, we will describe these
connections in more detail and see how they arise as the Berry
phase of certain states in the quantum mechanics.

\subsubsection*{3.1 $N=2$ and the Zeroth Hopf Map}

For the $N=2$ case, the ground state wavefunction undergoes
a discrete holonomy as the D0-branes orbit.  While the existence
of this phase follows on general grounds from the degenerate
nature of $H_F$ at the origin, it is instructive to review
explicitly how it occurs.

\para
The fermionic Hilbert space ${\cal H}_F$ consists of four states
$\vac$, $\Psi^\dagger_\alpha\vac$ and
$\Psi^\dagger_1\Psi^\dagger_2\vac$. The ground state may be
expanded as a linear combination of the middle sector,
\be |\Omega\rangle=\lambda_\alpha \Psi^\dagger_\alpha\vac\ee
with $\lambda_\alpha$ the negative eigenvector of $\hat{X}\cdot
\Gamma$.
We choose to work with the real basis of gamma matrices
$\Gamma^1=\sigma^1$ and $\Gamma^2=\sigma^3$ and introduce polar
coordinates
\be X^1=X\sin\theta\ \ \ ,\ \ \ X^2=X\cos\theta\ee
Then the groundstate eigenvector is
\be \vec{\lambda} =
\frac{1}{\sqrt{2+2\cos\theta}}\left(\begin{array}{c}-\sin\theta \\
1+ \cos\theta \end{array}\right) \label{turn}\ee
where we have resolved the square-root sign ambiguity in favour of
the positive. The evolution of the eigenvector as $\theta$ is
adiabatically varied is shown in the figure. We see that as
$\theta$ varies from 0 to $2\pi$, $\vec{\lambda}$ returns to
$-\vec{\lambda}$. This is the manifestation of the zeroth Hopf
map: $S^1\rightarrow S^1$.

\newcommand{\onefigurenocap}[1]{\begin{figure}[h]
         \begin{center}\leavevmode\epsfbox{#1.eps}\end{center}
         \end{figure}}
\newcommand{\onefigure}[2]{\begin{figure}[htbp]
         \begin{center}\leavevmode\epsfbox{#1.eps}\end{center}
         \caption{\small #2\label{#1}}
         \end{figure}}
\begin{figure}[htbp]
\begin{center}
\epsfxsize=2in\leavevmode\epsfbox{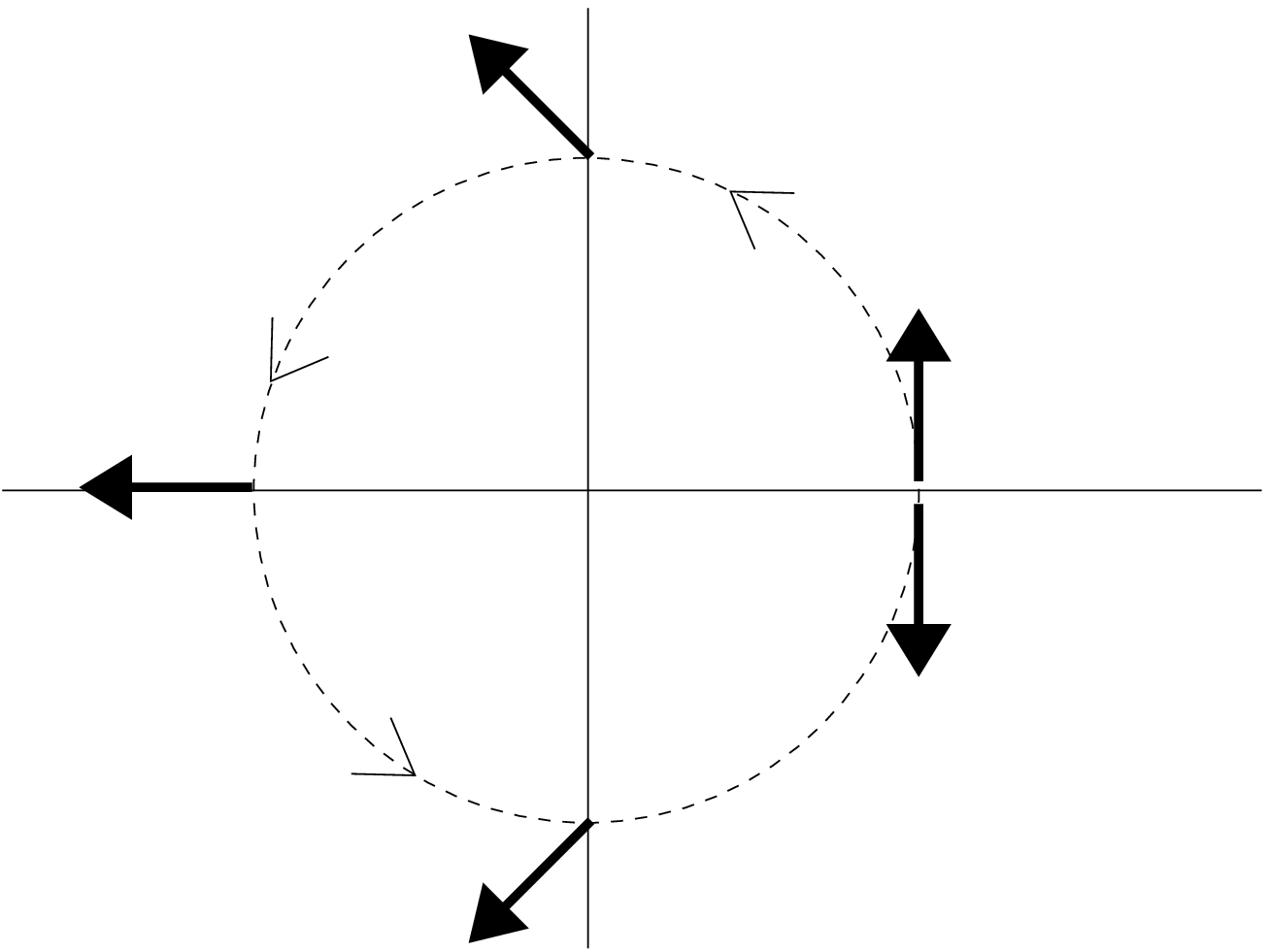}
\end{center}
{\small Figure 1: The ${\bf Z}_2$ holonomy as the eigenvector
encircles the degeneracy at the origin.}
\end{figure}

\para
This ${\bf Z}_2$ Berry phase is inherited by the ground state of
the D0-branes. After a single orbit, the wavefunction returns to
$|\,\Omega\rangle \mapsto -|\,\Omega\rangle$. Yet such an orbit is
equivalent to two exchanges of the D0-branes. Since the branes are
indistinguishable particles, upon a single exchange,
the D0-brane wavefunction picks up the phase $e^{\pm i\pi/2}=\pm
i$, where the $\pm$ sign depends on whether the exchange proceeds
clockwise or anti-clockwise. We learn that the D0-branes are
anyons \cite{anyon1,anyon2}.

\para
It is worthwhile re-deriving the fact that the D0-branes are
anyons by studying a single exchange, rather than a complete
orbit.  Let us start with the D0-branes separated by
$\hat{X}^i=(0,\hat{X})$, for which the ground state is given by
$|\Omega\rangle_+ = \Psi_2^\dagger\vac$. After an anti-clockwise
rotation to $\hat{X}^i=(0,-\hat{X})$, we see from \eqn{turn} that
the ground state adiabatically evolves into $|\Omega\rangle_- =
-\Psi_1^\dagger|0\rangle$. The ${\bf Z}_2$ Weyl symmetry ensures
that the Hilbert space constructed over $\hat{X}^i$ is physically
identified with the Hilbert space over $-\hat{X}^i$. Thus in order
to understand the phase picked up by the ground state, we need to
understand the map between the Hilbert spaces constructed over
$\pm\hat{X}^i$ induced by the Weyl group.

\para
The action of the ${\bf Z}_2$ Weyl group is given by,
\be &{\bf Z}_2:& \hat{X}^i \mapsto -\hat{X}^i\ \ \ ;\ \ \
z^i\mapsto -z^{\dagger i}\ \ \ ;\ \ \ \Psi\mapsto -\Psi^\dagger
\ee
From the latter of these actions, we see that the reference state
$\vac_+$ defined by \eqn{ref} at $\hat{X}^i$ is mapped to
$\Psi_1^\dagger \Psi_2^\dagger \vac_-$ at $-\hat{X}^i$, up to a
phase $\omega$,
\be &{\bf Z}_2:& \vac_+ \mapsto \omega \Psi_1^\dagger
\Psi_2^\dagger \vac_-\ee
The question is: what is $\omega$? We can answer this by following the
fate of the state $\Psi_1^\dagger\vac_+$ under two actions of the
Weyl group,
\be \Psi_1^\dagger\vac_+\mapsto -\omega\Psi_2^\dagger\vac_-\mapsto
-\omega^2 \Psi_1^\dagger\vac_+\ee
Insisting that the Weyl group action squares to the unit operator,
we learn that $\omega=\pm i$. We now use this knowledge to map the
state $|\Omega\rangle_-$ above back to an element of the Hilbert
space over $\hat{X}^i=(0,\hat{X})$. We see,
\be |\Omega\rangle_+ \stackrel{\tiny{\rm
adiabatic}}{\longrightarrow} |\Omega\rangle_- \stackrel{\tiny{\rm
{\bf Z}_2}}{\longrightarrow}\ \mp i|\Omega\rangle_+\ee
The upshot is that the ground state of the two D0-branes indeed
picks up the phase factor $\pm i$ upon exchange.

\para
Note that we have still to quantize the $U(1)\subset SU(2)$ Cartan
subalgebra. In particular, this includes the massless fermions
$(\psi_3)_\alpha$, which double the Hilbert space. The resulting
states have conjugate exchange statistics. If one state picks up a
phase $+i$ upon an anti-clockwise rotation, the other state picks
up a phase $-i$.

\subsubsection*{3.2 $N=4$ and the First Hopf Map}

We now turn to D0-branes moving in $d-1=3$ spatial dimensions.
As explained in Section 2, the Hilbert space is constructed by
acting with bosonic and fermionic creation operators on the
reference state $\vac$. Each of these creation operators is
accompanied by a projection operator: ${\cal P}$ for bosons,
defined in \eqn{bproj}, and $P_\pm$ for fermions, defined in
\eqn{fproj}. These two types of projection operators give rise to
two types of Berry holonomy:

\para
The projection operators ${\cal P}$ restrict the bosonic
excitations to lie tangent to the sphere $S^2$ at fixed
$|\hat{X}|$. As we saw above, this can be traced to the
implementation of the $SU(2)$ gauge symmetry and, from the string
theory perspective, is the familiar statement that stretched
strings have only transverse excitations. The net result is that as the
D0-branes orbit, excited states that include bosonic excitations
undergo a Berry holonomy as tangent vectors on the sphere $S^2$.
This is the same kind of
non-Abelian holonomy that was described in the original
paper of Wilczek and Zee \cite{wz}.

\para
Fermionic excitations are accompanied by the projection operators
$P_\pm$ \eqn{fproj}. To be concrete, let's focus on the operator
$P_-$. We introduce the normalized eigenvector
$P_-\lambda=\lambda$. Acting with the creation operator
$P_-\Psi^\dagger$ means creating the normalized state
$|\lambda\rangle =\lambda_\alpha\Psi_\alpha^\dagger\vac$. As
$\hat{X}^i$ varies adiabatically, this state undergoes a holonomy
described by the Berry connection,
\be A_i = i \langle \lambda | \frac{\partial}{\partial
\hat{X}^i}|\lambda\rangle\ee
which is the connection of the Hopf map \eqn{hopfcon}. An explicit
form of the connection requires a choice of gauge which, in this
context, means a chosen phase for the eigenvector $\lambda$. We
choose $\lambda_1\in \bR$, a choice which is valid everywhere
except along the positive $\hat{X}^3$ axis. In this gauge, the
Berry connection takes the familiar form of the Dirac monopole,
\be A_i =
\frac{-\hat{X}^j}{2\hat{X}(\hat{X}-\hat{X}^3)}\,\epsilon_{ij}
\qquad i = 1,2 \ \ \ \ ,\ \ \ \ A_3 =0 \label{dirac}\ee
The Hopf-Berry connection has first Chern class $-1$ over the
sphere $S^2$, meaning that the integral of the field strength
$F=dA$ yields,
\be c_1=\frac{1}{2\pi}\int_{S^2} F = -1\ee
Acting with the operator $P_+\Psi^\dagger$ results is a state
whose Berry connection has Chern class $+1$. The ground state in
the $N=4$ theory is given by
\be |\Omega\rangle = N^\prime(P_-\
\Psi^\dagger)(P_+\tilde{\Psi}^\dagger)\vac\equiv (\lambda_\alpha
\Psi_\alpha^\dagger)(\tilde{\lambda}_\beta\tilde{\Psi}_\beta^\dagger)\vac\ee
where $N^\prime$ is a normalization factor, while $\lambda$
($\tilde{\lambda}$) is the normalized non-zero eigenvector of
$P_\mp$. The presence of the two, opposite, projection operators
ensures that the ground state does not pick up a Berry phase as
the D0-branes orbit.

\para
Excited states do pick up a Berry phase, given by the sum of the
phases associated to the relevant projection operators. For
example, a degenerate pair of states obeying Gauss' law
is given
by
\be |\phi_-\rangle = (\lambda_\alpha\Psi_\alpha^\dagger)
(\lambda_\alpha\tilde{\Psi}_\alpha^\dagger)\vac \ \ \ ,\ \ \
|\phi_+\rangle = (\tilde{\lambda}_\alpha\Psi_\alpha^\dagger)
(\tilde{\lambda}_\alpha\tilde{\Psi}_\alpha^\dagger)\vac \ee
These states have energy $E=E_B+E_F=\hat{X}/g^2$ and describe two
strings attached to the D0-branes. (They are part of a triplet of
excited spin 1 states).
%
%
From the discussion above, we see that these states pick up a
Berry phase arising from a magnetic monopole of charge $q=\pm2$.
Physically, this means that the D0-branes orbit as charged
particles as if in the presence of a magnetic monopole fixed at
their centre. In a semi-classical analysis, the orbits are no
longer restricted to lie on a plane, but rather lie on a cone with
opening angle $\cos\theta = -q/(2J+q)$, where $J$ is angular
momentum of the spinning D0-branes. The energy $E$ these rotating
states scales as $E^3\sim g^2 J(J+q)$.

\subsubsection*{3.3 $N=8$ and the Quaternionic Hopf Map}

The discussion for the $N=8$ theory is very similar to the $N=4$
theory above. The unique ground state of the system \eqn{vac2}
does not undergo a Berry phase.
%
%
Excited states do. The typical state undergoes a non-Abelian
holonomy arising from the sum of bosonic and fermionic Berry
connections. Once again, bosonic excitations give rise to a
holonomy in which states transform as tangent vectors on $S^5$.
More interesting for the present discussion are the fermionic
excitations. The projection operator $P_-$ now has a pair of
orthonormal eigenvectors: $P_-\lambda_a=\lambda_a$, $a=1,2$. This
means that the resulting Berry holonomy of the states
$|\lambda_a\rangle= \lambda_{a\alpha}\Psi^\dagger_\alpha\vac$ is
described by a $U(2)$ connection,
\be (A_i)_{ab} = i\langle \lambda_a |\frac{\partial}{\partial
\hat{X}^i}|\lambda_b\rangle \ee
Explicit computation shows that this is actually an $SU(2)$
connection, known as the Yang-monopole \cite{yang}. A suitable
choice of gamma matrices and gauge, can be found in \cite{second}.
The resulting connection is,
\be (A_i)_{ab} =
\frac{-\hat{X}^j}{2\hat{X}(\hat{X}-\hat{X}^5)}\,\eta^m_{ij}\,\sigma^{m}_{ab}
\qquad i = 1,2,3,4 \ \ \ \ ,\ \ \ \ A_5 =0 \label{yang}\ee
where $\eta_{\mu\nu}^m$ are the self-dual $4\times 4$ 't Hooft
matrices and $\sigma_{ab}$ are the Pauli matrices. The Yang
monopole is perhaps more familiar when viewed as a connection
restricted to $S^4$, where it is simply the $SO(5)$ invariant
instanton satisfying $F=-{}^\star F$ with second Chern class,
\be c_2=\frac{1}{8\pi^2}\int_{S^4}\,{\rm tr}(F\wedge F)=-1 \ee
where the non-Abelian field strength is defined by
$F_{ij}=\partial_iA_j-\partial_jA_i-i[A_i,A_j]$. States
constructed from the projection operators $P_+$ undergo a Berry
holonomy associated with a Yang-monopole of Chern class $c_2=+1$.


\subsubsection*{3.4 $N=16$ and the Last Hopf Map}

Details for the theory with $N=16$ supercharges are again similar
to those above: the vacuum \eqn{vac1} does not pick up a Berry
connection, while bosonic excitations transform as tangent vectors
on $S^8$. The fermionic excitations are associated to projection
operators $P_\pm$, which are now $16\times 16$ real matrices. We
once again define the $8$ orthonormal, real eigenvectors
$P_-\lambda_a=\lambda_a$, with $a=1,\ldots,8$. States in the
quantum mechanics involving fermions undergo a holonomy arising
from the $SO(8)$ Berry connection,
\be (A_i)_{ab} = i\langle \lambda_b |\frac{\partial}{\partial
\hat{X}^i}|\lambda_a\rangle\ee
with $|\lambda_a\rangle =
\lambda_{a\alpha}\Psi_\alpha^\dagger\vac$. We refer to the
resulting connection as the $SO(8)$ octonionic monopole. It was
constructed in \cite{grossman,tchrak}. A simple expression for the
connection can be found in \cite{bern}
\be (A_i)_{ab} =
\frac{-\hat{X}^j}{2\hat{X}(\hat{X}-\hat{X}^9)}\,\Sigma_{ij} \qquad
i = 1,\ldots,8 \ \ \ \ ,\ \ \ \ A_9 =0 \label{so8}\ee
where $\Sigma_{ij}$ are the 28 generators of $SO(8)$ Lie algebra,
defined in terms of the $\Gamma$ matrices. An explicit form can be
given if we choose a suitable representation for the generators of
the octonions $e_i$, $i=1\ldots 8$ in \eqn{gam}, in terms of
$8\times 8$ real matrices. We choose the unit matrix to correspond
to $e_8=1$. Then $\Sigma_{ij} = -\ft12 [e_i,e_j]$, for
$i=1,\ldots,7$ and $\Sigma_{i8}=e_i$ \cite{bern}. The non-Abelian
field strength restricted to $S^8$ satisfies the generalized
self-duality condition  $F\wedge F = -{}^\star F\wedge F$, with
\be \frac{1}{4!(2\pi)^4}\int_{S^8}{\rm tr}(F\wedge F\wedge
F\wedge F) = -1\ee
%
%
%
As in previous cases, states associated to $P_+$ undergo a Berry
holonomy with $c_4=1$.

\para
It is intriguing to see the octonionic monopole appearing in the
Matrix theory for D0-branes in this fashion, although its physical
significance in M-theory remains obscure. Nonetheless, it is
tempting to speculate. Firstly, recall that there is a precedent
for the appearance of Berry's phase in Matrix theory: the membrane
feels the magnetic field of the five-brane through the appearance
of a Berry connection \cite{dberk}. In this case, the abstract
Berry connection is recast as a physical magnetic field. It would
be very interesting if a similar M-theory interpretation could be
given in the present case. We note in passing that a physical
octonionic monopole is conjectured to act as the end-point for an
open heterotic string \cite{pol}.

\subsection*{Acknowledgments}

We would like to thank Freddy Cachazo, Joe Polchinski, Savdeep
Sethi and Paul Townsend for useful discussions. J.S. gratefully
acknowledges the University of Chicago, Harvard and the Perimeter
Institute for hospitality and travel support while this paper was
being written up. D.T. thanks the Isaac Newton Institute for their
kind hospitality while this work was undertaken. C.P. is supported
by an EPSRC studentship. J.S. is supported by a research
fellowship from Trinity College, Cambridge. D.T. is supported by
the Royal Society.

\end{document}